\documentclass{llncs}

\usepackage{subfig}
\usepackage{times}
\usepackage{graphicx}

\newtheorem{defn}{Definition}
\begin{document}
\title{Coordination of Mobile Mules via Facility Location Strategies}

	\author{Danny Hermelin \and Michael Segal  \and Harel Yedidsion  }
		
	\institute	{ Ben-Gurion University of the Negev, Beer-Sheva, Israel,\\	
	\email	{hermelin, segal, yedidsio @bgu.ac.il}		 }

\maketitle

\begin{abstract}
In this paper, we study the problem of wireless sensor network (WSN) maintenance using mobile entities called \emph{mules}. The mules are deployed in the area of the WSN in such a way that would minimize the time it takes them to reach a failed sensor and fix it. The mules must constantly optimize their collective deployment to account for occupied mules.

The objective is to define the optimal deployment and task allocation strategy for the mules, so that the sensors' downtime and the mules' traveling distance are minimized.

Our solutions are inspired by research in the field of computational geometry and the design of our algorithms is based on state of the art approximation algorithms for the classical problem of facility location.

Our empirical results demonstrate how cooperation enhances the team's performance, and indicate that a combination of $k$-Median based deployment with closest-available task allocation provides the best results in terms of minimizing the sensors' downtime but is inefficient in terms of the mules' travel distance. A $k$-Centroid based deployment produces good results in both criteria.
\end{abstract}

\section{Introduction}
\noindent Wireless Sensor Networks (WSN) have recently become a
prevalent technology used in a wide range of environmental
monitoring applications such as temperature, pollution and wildlife
monitoring. Typically a WSN is composed of a large number of sensor nodes ($n$)
coupled with short range radio transceivers. The sensors transfer
their sensed data to a central hub via multi-hop communication. A
communication tree is formed based on physical proximity and must be
maintained through technical failures such as battery drainage or
memory overload. The root of such a tree is usually a special node
having significant power and communication abilities (for example,
it is able to send an alert message to the control center which is
far away from the actual sensors' location). In case some sensor in the
communication tree fails, it not only stops monitoring its
environment but, it might also disconnect the communication from
other parts of the network.

We study the use of mobile agents called {\em mules} which have
the ability to reach failed nodes, fix them and temporarily replace
their role in the task of data collection and transfer. The mules
are used to maintain and improve the networks resiliency and
reliability. 

We aim to optimize the mules' deployment and to design the
cooperation protocol between them as to minimize the duration of
failures and the mules' travel distance. The decision on which exact objective function to measure involves some conflicting considerations.

By limiting the maximal downtime that any sensor may experience we can make sure that there is no loss of data for longer than a certain period of time. 
This guarantee is important for WSNs where data is sensed or transmitted periodically. On the other hand, minimizing the average downtime minimizes data loss in WSNs where sensors constantly sense and transmit data.

As for the mules' movement, since the mules are battery operated, it is important to limit the maximal movement by any mule to prevent its total battery depletion. Nonetheless, minimizing the mules' average travel distance would extend the total lifetime of the team as a whole and would enable them to fix more failures. As a consequence, we try to minimize both objectives (i.e., downtime and movement) under the two criteria (i.e., average and max) while focusing on minimizing the average downtime.

We propose algorithms that differ in their positioning methods and their suggested level of cooperation between mules. The contribution of our work lies in the novel use of known facility location approximation algorithms for solving the mule team problem.
    \vspace{-10pt}
\begin{defn}
    \label{KCenter}
    The $k$-Center problem is defined as follows. Given a set of $n$ points $S$ and an integer $k$, find a set $S^{\prime}$ of $k$ points for which the largest euclidean distance of any point in $S$ to its closest point in $S^{\prime}$ is minimum.
\end{defn}
 \vspace{-10pt}
\begin{defn}
    \label{KMedian}
    The $k$-Median problem is defined as follows. Given set of $n$ points $S$ and an integer $k$, find a set $S^{\prime}$ of $k$ points for which the sum of euclidean distances of any point in $S$ to its closest point in $S^{\prime}$ is minimum.
\end{defn}
 \vspace{-10pt}
\begin{defn}
    \label{Centroid}
    The centroid of a set of points $S$ is the arithmetic mean of their coordinates.
\end{defn}
 \vspace{-10pt}
\begin{defn}
    \label{VoronoiCell}
    A Voronoi diagram is a partitioning of a plane into regions based on distance to a given set of points $S$. For each point in $S$ there is a corresponding region consisting of all points in the plane closer to that point than to any other point in $S$. These regions are called Voronoi cells.
\end{defn}
 \vspace{-10pt}
\subsection{Related Work}
Coordinating a team of mobile agents to perform tasks in a dynamic environment is a fundamental problem in AI that has received much attention from researchers \cite{StoneKKR10,GerkeyM04,urra2009using}. Some of the popular methods that have been used to tackle this problem include:
game theory \cite{Tambe11,JiangPQST13}, machine learning \cite{Krause08,WangD08},  multi-agent path planning and scheduling \cite{GenterS15,PitaJOPTWPK08}, distributed constraint optimization \cite{FarinelliRJ13,ZivanYOGS15}, economic market based approaches and auctions \cite{KoenigKT10,McIntireNG16}, virtual potential fields \cite{PoduriS04} and probabilistic swarm behavior \cite{KonurDF12}.

We aim to further contribute to the study of multi-agent coordination by investigating the integration of known facility location techniques into the design of the algorithms which are meant to solve the mule team problem.

The term MULE (Mobile Ubiquitous LAN Extensions) was coined in \cite{ShahRJB03} to refer to mobile agents capable of short-range wireless communication that can exchange data with a nearby sensor.\footnote{We use the term mules and agents interchangeably.}
In the field of WSN, mobile elements have been proposed to improve maintenance, data collection, connectivity and energy efficiency \cite{DiFranDA11}.
Crowcroft et al. \cite{CrowcroftLS16} define the $(\alpha, \beta)$-Mule problem, where $\alpha$ is the number of simultaneous node failures and $\beta$ is the number of traveling mules. Unlike our work where the topology of the network is given, their aim is to define the topology of the network in order to minimize the mules' tours.
In \cite{LevinES14}, Levin et al. study the tradeoff between the mules' traveling distance and the amount of information uncertainty caused by not visiting a subset of nodes by the mules.
The authors of \cite{RuiLM16} utilize autonomous mobile base stations (MBSs) to automatically construct new routes to recover disconnected infrastructure, while in \cite{AnandZM09} mobile backbone nodes (MBNs) are controlled in order to maintain network connectivity while minimizing
the number of MBNs that are actually deployed.
In \cite{TekdasILT09}, the $k$-Traveling Salesman Problem (TSP) approach is used to plan the data collection routs of $k$ mules. TSP with neighborhood was applied for the same purpose in \cite{Kim12}.

The $k$-Server problem proposed by \cite{Manasse1990} and researched by many others is closely related to our problem. However, the major difference is that the objective of the $k$-Server problem is to minimize the total traveling distance of the servers while in our problem this is a secondary objective.
Our main interest is finding an optimal deployment scheme for the mules as to minimize the downtime of failed sensors. 
Another similar strand of work relates to the ambulance redeployment problem \cite{maxwell2010approximate,yue2012efficient}. The difference from our model is that the set of deployment bases is predefined and finite while in our case agents can be placed anywhere in the environment.


The static version of a single iteration of our problem relates to the $k$-Center and $k$-Median problems (see Definitions \ref{KCenter} and \ref{KMedian} respectively). Since both of these problems are NP-Hard, we integrate into our algorithms known greedy heuristics for these problems that guarantee some approximation ratio of the optimal solution. These solutions run in polynomial time and thus are suited for practical applications that require quick responses. Specifically, Gonzalez \cite{Gonzalez85} proposed the Farthest-First (FF) algorithm that provides a 2-approximation for the  $k$-Center problem in $O(n)$ time. In \cite{ChrobakKY05} it is shown that the reverse greedy algorithm (RGreedy) guarantees at most an approximation ratio of $O(\log n)$ in $O(n^2 \log n)$ time for the $k$-Median problem.

\subsection{Problem Definition}
Formally, the mule team problem is defined as follows: $V$ is the
set of $n$ wireless sensor nodes embedded in the Euclidean plane.
Let $M$ be a set of $m$ mobile agents (mules) that can
travel anywhere in the plain and fix sensors that experience
technical failures. The sensors are subject to a set of failures
$F$. Each failure occurs at a certain node, at a certain time and
has a predefined failure duration $F_d$, which is the time it takes
to fix a node from the moment a mule has reached it. The term
"downtime" of a node $v$, denoted $v_d$, refers to the time from
when $v$ failed until a mule reaches it.
Mule $m$'s travel distance is denoted $m_t$
The mules are immediately
aware of any failure and can communicate with each other. Once a
mule is engaged in fixing a failed node $v$, experiencing failure $f$, it is unable to attend
to any other tasks for the time it takes to travel to $v$ plus
$f$'s specified fail duration, $f_d$. It is assumed that the time to
fix a failure is greater than the average time between consecutive
failures and much greater than the average time it takes a mule to
move to a failure.

The \textbf{goal} is to find a continuous deployment strategy and a
cooperation method for the mules, which minimize two opposing
objectives. The primary objective is to minimize the nodes' downtime
and the secondary objective is to minimize the mules' traveling
distance. These objectives are measured according to two criteria,
namely average and max, while our focus is on the average criterion,
we also monitor the max criterion. There exists a trade-off between
these two objectives since minimizing downtime requires the mules to
redeploy after every failure thus increasing their travel distance.
The \textbf{challenge} is to design an algorithm that would produce
the best results on both objectives and according to both criteria.

The first objective is to minimize the nodes' average downtime and is formalized as: $\min(\sum_{v \in V}v_d/|V|)$. 
The second objective is to minimize the mules' average travel distance:
$\min( \sum_{m \in M} m_t/|M|)$.  
The same objectives under the max criterion are: 
minimizing the nodes' maximal downtime, $\min(\max_{v \in V}v_d)$, and minimizing the mules' maximal traveling distance, $\min( \max_{m \in M} m_t)$. 

\section{Algorithms}
\subsection{Facility Location Strategies}
The facility location problem is a well studied problem in computer science and operations research \cite{ReVelleE05}. It has many variations which primarily deal with optimally placing $k$ facilities to service $n$ given cities. Two classical variants of this problem which are closely related to our problem are the $k$-Center and $k$-Median problems, both proven NP-hard problems \cite{KarivH179,KarivH279}. Consequently, there is a significant body of work that deals with approximation algorithms for these problems.

An optimal solution to the $k$-Center problem minimizes the maximal distance of any city to its closest facility and thus, any algorithm that provides a good solution for this problem can be useful in the design of an algorithm that would minimize the mules' maximal movement and the sensors' maximal downtime in the mule team problem.

The FF algorithm proposed in \cite{Gonzalez85} provides a 2-approximation for the $k$-Center problem in $O(n)$ time. The algorithm greedily selects $k$ points in the following way. The first point is selected arbitrarily and each successive point is chosen out of the $n$ nodes as far as possible from the set of previously selected points.

An optimal solution to the $k$-Median problem minimizes the sum of distances of all the cities from their closest facility and thus, any algorithm that provides a good solution for this problem can be useful in the design of an algorithm that would minimize the mules' average movement and the sensors' average downtime in the mule team problem.

The Reverse Greedy algorithm (RGreedy) proposed in \cite{ChrobakKY05} to solve the $k$-Median problem, works as follows.
It starts by placing facilities on all nodes. At each step, it removes a facility to minimize the total distance to the remaining facilities. It stops when $k$ facilities remain. It runs in $O(n^2\log n)$ time.

Finally, in \cite{MilyeykovskiSK15} it was proven that a centroid of
a set of points $P$ provides a 2-approximation for the 1-median of
$P$. Table 1 summarizes these findings.

It should be mentioned that there are additional approaches to deal
with variants of $k$-Center and $k$-Median problems. So-called
$\varepsilon$-nets~\cite{MustafaR10} and Linear Programming
relaxation~\cite{CharikarGTS02} are just few examples. However,
these approaches do not allow distributed implementation which is
essential to make our solutions feasible for real life deployments.

\begin{center}
    \begin{table}[h]
    	    \vspace{-20pt}
        \renewcommand{\arraystretch}{1.3}
        \centering
        \begin{tabular}{|c|c|c|c|}
            \hline Name & Reference & Performance Bounds \\
            \hline FF & Gonzalez & 2-Apx. for $k$-Center \\
            \hline RGreedy & Chrobak et al. & $\log n$-Apx. for $k$-Median \\
            \hline Centroid & Milyeykovski et al. & 2-Apx. for $1$-Median \\
            \hline
        \end{tabular}
       \vspace{10pt}
        \caption{Summary of facility location approximation algorithms}
       \vspace{-40pt}
        \label{tab:Table_apx}
    \end{table}
\end{center}

\subsection{Proposed Algorithms}
The proposed algorithms differ in their approach to four main traits, i.e., the mules' initial deployment, task allocation, continuous redeployment and cooperation methods. Each trait can be implemented in several ways.
\begin{enumerate}
    \item The mules' initial deployment:
    \begin{itemize}
        \item Grid - The mules are uniformly distributed in the area of the nodes.
        \item Farthest-First - The mules are deployed according to the FF algorithm which approximates the $k$-Center problem.
        \item Reverse-Greedy - The mules are deployed according to the RGreedy algorithm which approximates the $k$-Median problem.
        \item Centroid-Adjustment - Each of the above methods can be combined with an additional repositioning stage of centroid adjustment where each mule moves to the centroid position of its closest nodes. This process is performed iteratively until convergence.

    \end{itemize}

        \item The mules' cooperation method:
        \begin{itemize}
            \item No cooperation - Each mule is in charge of the nodes in its Voronoi cell according to the initial deployment. This allocation is static and does not change as the mules move.
            \item Cooperation - Here there is no strict node-to-mule allocation and every mule can fix any node even if it lies in another mule's Voronoi cell.

        \end{itemize}

    \item The mules' task allocation strategy (Only applicable for cooperative algorithms):
    \begin{itemize}
        \item Closest - Send the closest mule to each failure thus minimizing average travel distance.
        \item Closest-Available - Send the closest available mule thus minimizing downtime.
        \item Closest-Least Traveled - Send the closest mule whose total travel distance after tending to the current failure is the  lowest, thus minimizing the maximal travel distance.

    \end{itemize}

        \item The mules' redeployment:
        \begin{itemize}
            \item No redeployment - This case has two options, either the mule that moves simply stays in its new position to minimize travel distance or it returns to its initial position to return to a deployments that was calculated to offer good reaction times and minimize downtime.
            \item Farthest-First - The available mules are redeployed according to the FF algorithm and the occupied mules are disregarded. After recalculating the new positions, the closest mule is sent to every new location as to minimize their traveling distance during redeployment.
            \item Reverse-Greedy - The mules are redeployed according to the RGreedy algorithm yet only $(k-b)$ medians are calculated out of $(n-b)$ node locations where $b$ is the number of busy, occupied mules. Occupied mules and nodes that are being fixed are disregarded in the RGreedy calculation. After recalculating the new positions, the closest mule is sent to every new location as to minimize their traveling distance during redeployment.

            \item Centroid-Adjustment - Unoccupied mules perform centroid adjustment by moving to the centroid position of their closest nodes while disregarding occupied mules.

        \end{itemize}

\end{enumerate}

The redeployment stage poses another interesting problem of mule's
reassignment, i.e. which mule to assign to which location as to
minimize the total traveling distance of mules. 
This problem is
similar to the Minimum Weight Bipartite Matching Problem which can be solved optimally in $O(n^3)$ time, where $n$ is the number of assignments, using the Hungarian Algorithm, \cite{Munkres57}. We implemented this algorithm to determine the assignments of mules to new locations of the redeployment.

The different combinations of traits produce a large number of possible algorithms. Although we tested many algorithms in a wide variety of settings we limit our analysis to the few algorithms that yielded the best results and in addition, provide clear insights on the performance of the facility location techniques.
Here are the proposed algorithms:

\begin{itemize}

\item  \textbf{Basic Grid Algorithm} - This algorithm is designed to be a baseline algorithm to which we will compare the others. In our preliminary testing it provided the best results out of the algorithms that do not perform any redeployment. It uses uniform grid placement for the initial mules' deployment. The closest-available mule is assigned to a failure. No redeployment is performed.

\item  \textbf{No Cooperation Algorithm} - This algorithm is designed to compare the performance of the non-cooperative approach to the cooperative one. It uses uniform grid placement for the initial mules' deployment. Each mule is assigned the nodes in its Voronoi cell. This assignment is constant and there is no cooperation. This means that if a failure occurs in a Voronoi cell of an occupied mule, none of its neighbors would help out even if they are available. No redeployment is performed.

\item  \textbf{ $\mathbf{k}$-Center Algorithm} - This algorithm uses FF for the initial mules' deployment and for their redeployment. Mules cooperate and the closest-available allocation is used.

\item  \textbf{$\mathbf{k}$-Median Algorithm} - This algorithm uses RGreedy for the initial mules' deployment and for their redeployment. Mules cooperate and the closest-available allocation is used.

\item  \textbf{$\mathbf{k}$-Centroid Algorithm} - The initial deployment is done using FF but is immediately followed by a procedure called centroid-adjustment, where each mule moves to the centroid of the nodes in its Voronoi cell. Centroid-adjustment is also used to redeploy the mules after every movement. Here too, the closest-available mule is sent to any failure.

\item  \textbf{Local Search Algorithm} - While approximation algorithms can provide certain theoretical guarantees, there is no guarantee that they perform as good as local search methods in practice. For this reason we also implemented a local search algorithm to evaluate the overall performance of our proposed algorithms which are based on approximation algorithms. 

One of the difficulties in applying local search is the fact that the number of steps to reach a local optimum could be exponential. Since we are dealing with low polynomial time solutions, it would not be fair to compare between both approaches. We deal with this issue is by using a limited time frame for search and an anytime mechanism for maintaining the best achieved state during the search process. 
The mules' deployment strives to achieve the $k$-Median criterion.
The search process is performed in a distributed concurrent manner as each agent incrementally moves to a nearby position that minimizes the sum of distances from it to the nodes closest to it.
The number of iterations is limited to a constant, the number of nodes $n$, or until convergence thus maintaining $O(n)$ runtime.
Here too, the closest-available mule is sent to any failure.

\end{itemize}

\section{Experimental Evaluation}
In order to compare the performance of the different algorithms, we developed a software simulator, representing a WSN with failures and a team of mobile mules. The area of the simulated problem is an $X$ over $Y$ plane. Any number of nodes ($N$), and mules ($M$) can be positioned in the area. Within the total duration of the experiment ($E_t$), we can randomly induce any number of failures ($F$) on the nodes either in uniform or in non-uniform distribution. The nodes which fail are randomly chosen from $N$ and the start time of each failure is randomly chosen from $(0,E_t)$. For each experimental setting we test several values of failure durations ($F_d$), which is the time it takes to fix a failure from the time a mule has reached it.

In each experiment the specific values for $N, M, E_t, F, F_d$  are chosen differently to demonstrate the algorithms' behavior in different scenarios. We ran several tests to find a combination of parameters that gives a good separation in the algorithms’ performance and that adhere to natural assumptions of such a practical system.
The parameter setting in general follows these assumptions:
$N > M$, and 
Failure duration $\gg$  Average time between failures $\gg$ Average travel time to reach a failed node.
Each reported result represents an average of 50 random experiments. In each experiment the initial mule locations, node locations, failure location and start times are randomly selected. We used the same set of random seeds so that each algorithm is presented with the same 50 randomly generated problems.
To analyze the statistical significance of the results, we performed T-Tests to validate the difference between the algorithms' performances. We state verbally in the text whether the difference between the results are statistically significant (i.e. p-value $<0.05$) or not. For sake of readability, we refrain from adding error bars to the figures.


In the following subsections we analyze specific representative cases.

\subsection{Comparing Cooperative vs. Non-Cooperative Algorithms}
\label{Cooperation}
In this subsection we analyze the differences between the performance of non-cooperative algorithms where the node-to-mule assignment is static, to cooperative algorithms where any mule can attend to any failure even if it is not the closest mule to this failure. To this end we use the No-Cooperation algorithm to represent non cooperative behavior. It uses grid initial deployment and static task allocation, the mules do not return to their initial position as this strategy proved better than returning to the initial position on both objectives.

The experimental setting included problems with 10 mules and 100 nodes that were randomly deployed in a $X=100$ over $Y=100$ area. 100 failures were generated with $F_d={0,100,...,1000}$ and $E_t=10000$. Failure distribution is uniform i.e, each sensor has the same probability of failing.
\begin{figure}
	  \vspace{-10pt}
	\centering
	\begin{minipage}{0.45\textwidth}
		
		\centering
		\includegraphics[width=6cm]{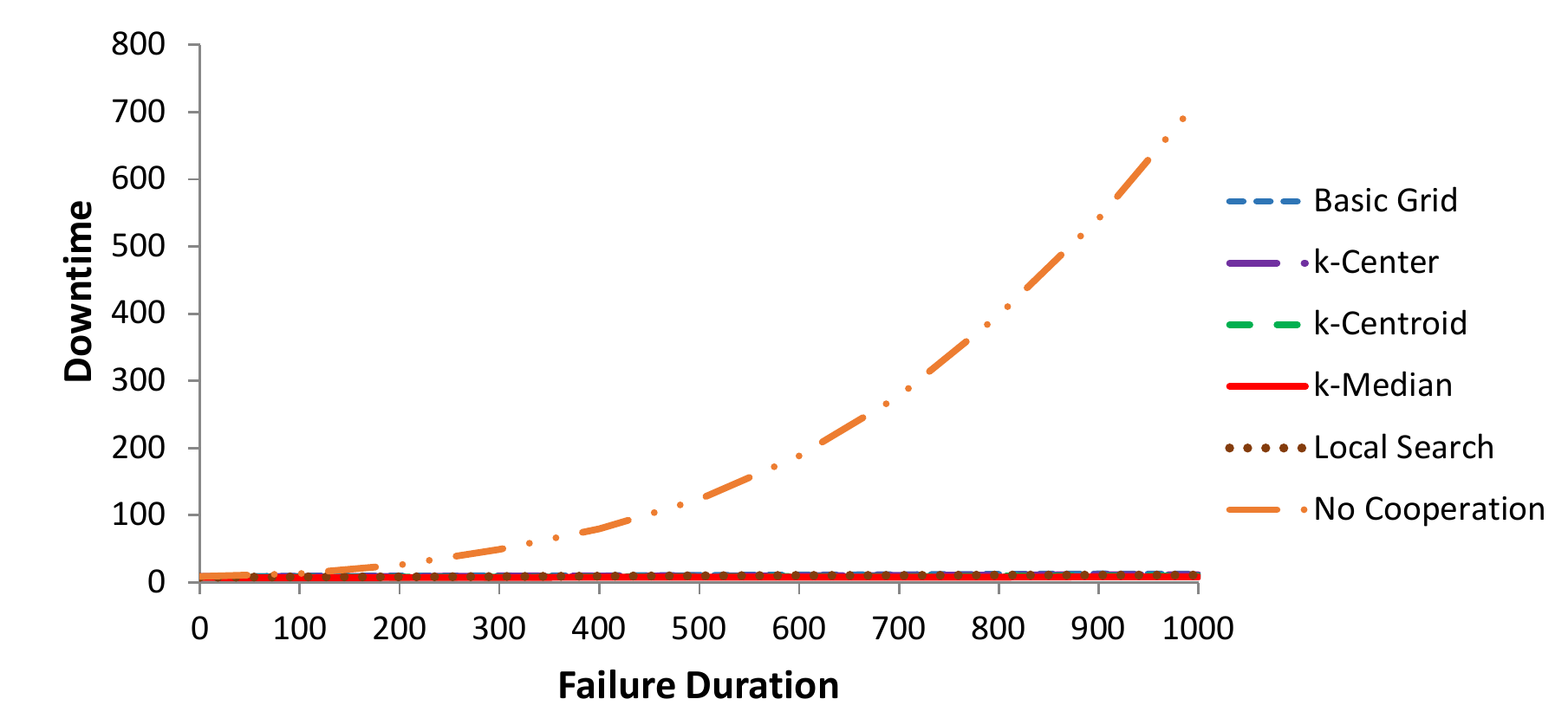}
		\caption{Comparing cooperative vs. non-cooperative algorithms.}
		\label{Fig:FigureCoop}
		
	\end{minipage}\hfill
	\begin{minipage}{0.45\textwidth}
		\centering
		\includegraphics[width=6cm]{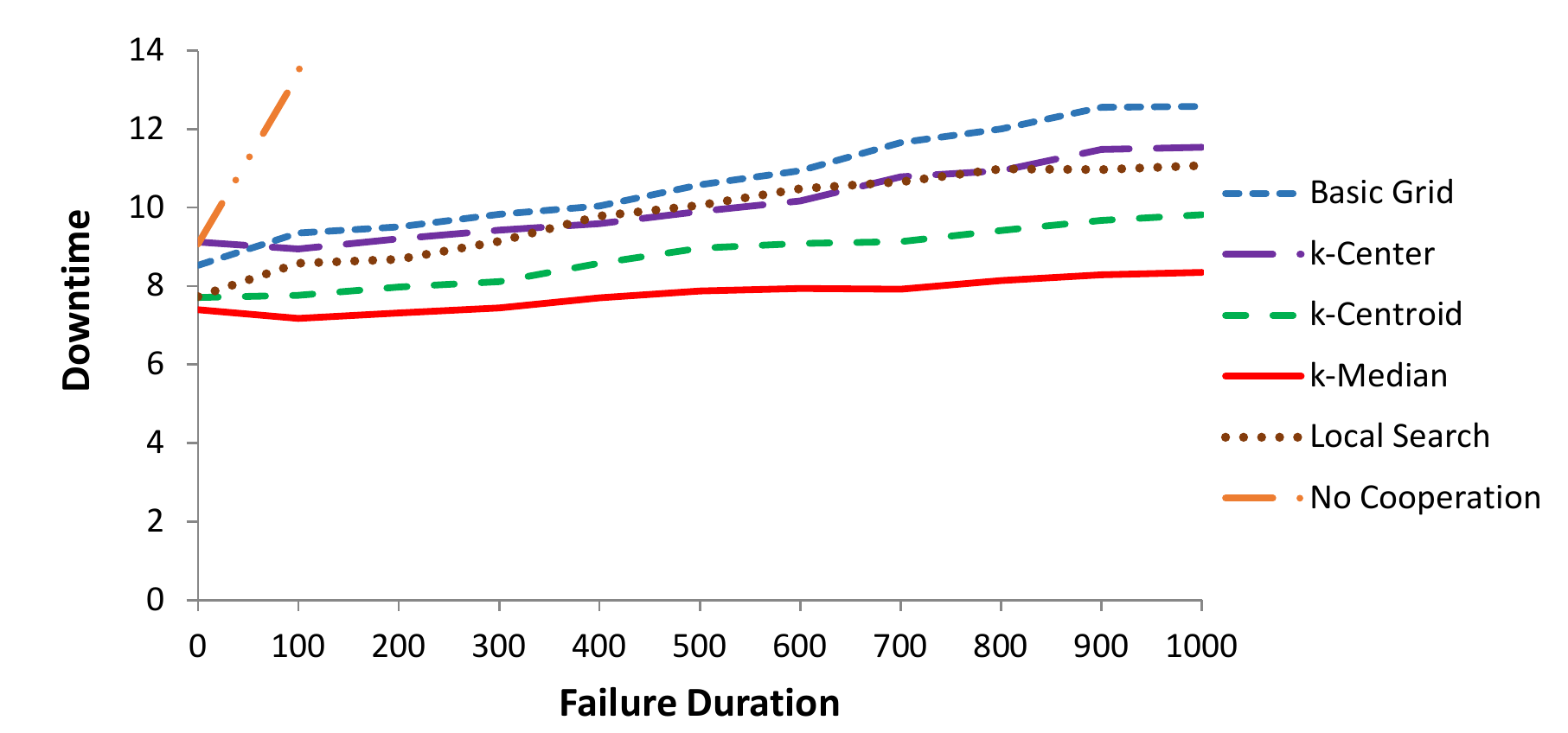}
		\caption{A closer look.}
		`	 \label{Fig:FigureCoop1}
	\end{minipage}\hfill
	\vspace{-15pt}
\end{figure}

Figures~\ref{Fig:FigureCoop} and ~\ref{Fig:FigureCoop1} present a comparison of the average downtime per failure between cooperative and non-cooperative algorithms. It is evident that as failure durations increase, the non-cooperative algorithm's performance worsens. The $k$-Median algorithm performs significantly better than all other algorithms.

\subsection{Comparison of Cooperative Algorithms}
\label{Sec:experimental1}

This subsection presents a comparison between the three facility location inspired algorithms (i.e., $k$-Center, $k$-Median, $k$-Centroid) and, following the conclusions derived in the previous subsection, we do not compare them to a non cooperative algorithm but instead to the best performing cooperative algorithm that does not use any redeployment i.e., the Basic Grid algorithm, and to the Local Search algorithm.

The experimental setting uses the following parameter values: 10 mules, 100 nodes, $X=100$ over $Y=100$ area, 10 failures with $F_d={0,1000,...,10000}$ and $E_t=10000$. Failure distribution is uniform.
\begin{figure}
	\centering
    \begin{minipage}{0.45\textwidth}

		\centering
        \includegraphics[width=6cm]{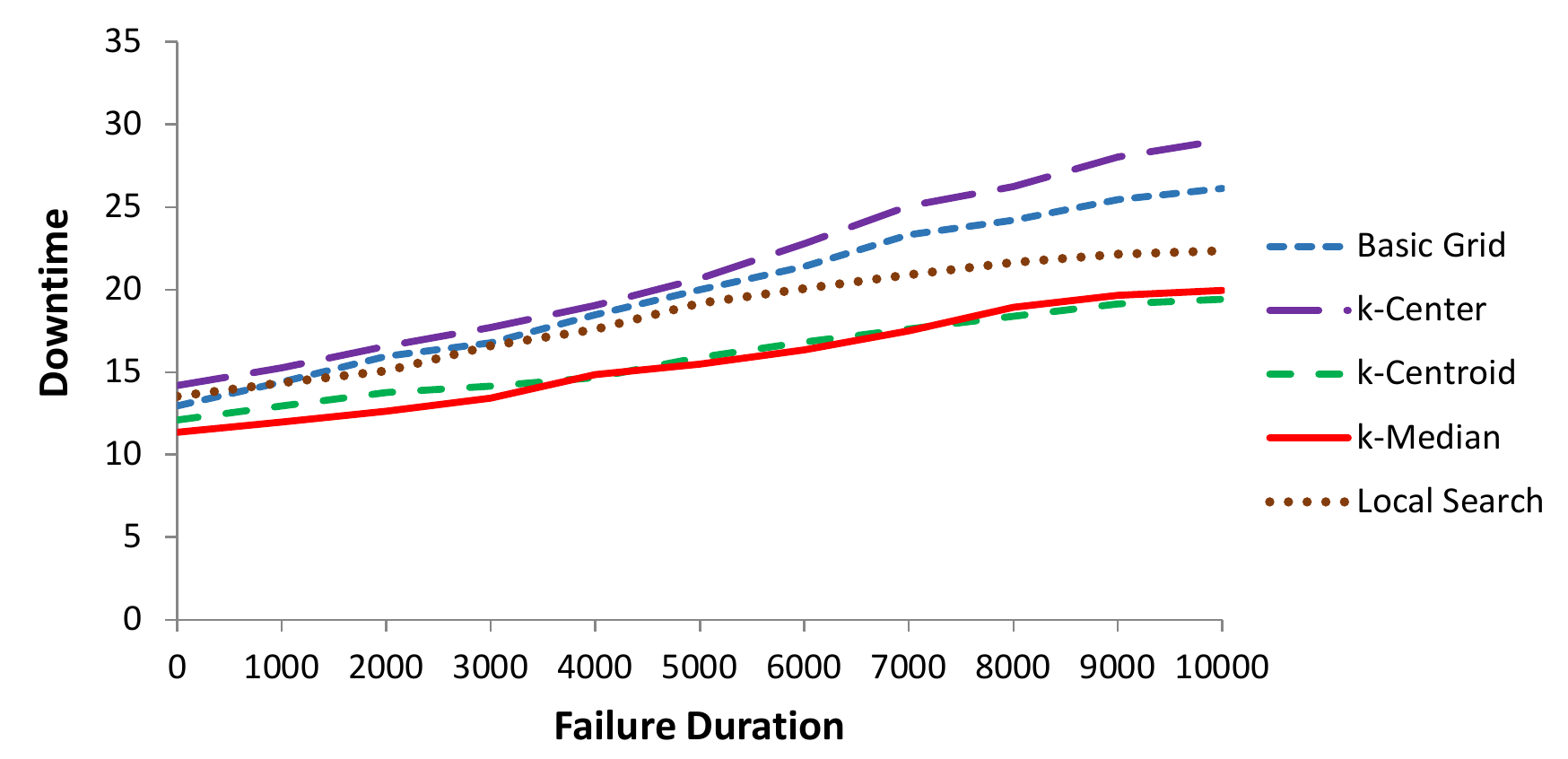}
       \caption{Average downtime per failure as a factor of increasing failure durations.}
        \label{Fig:Figure1ADT}

 \end{minipage}\hfill
 \begin{minipage}{0.45\textwidth}
		\centering
    \includegraphics[width=6cm]{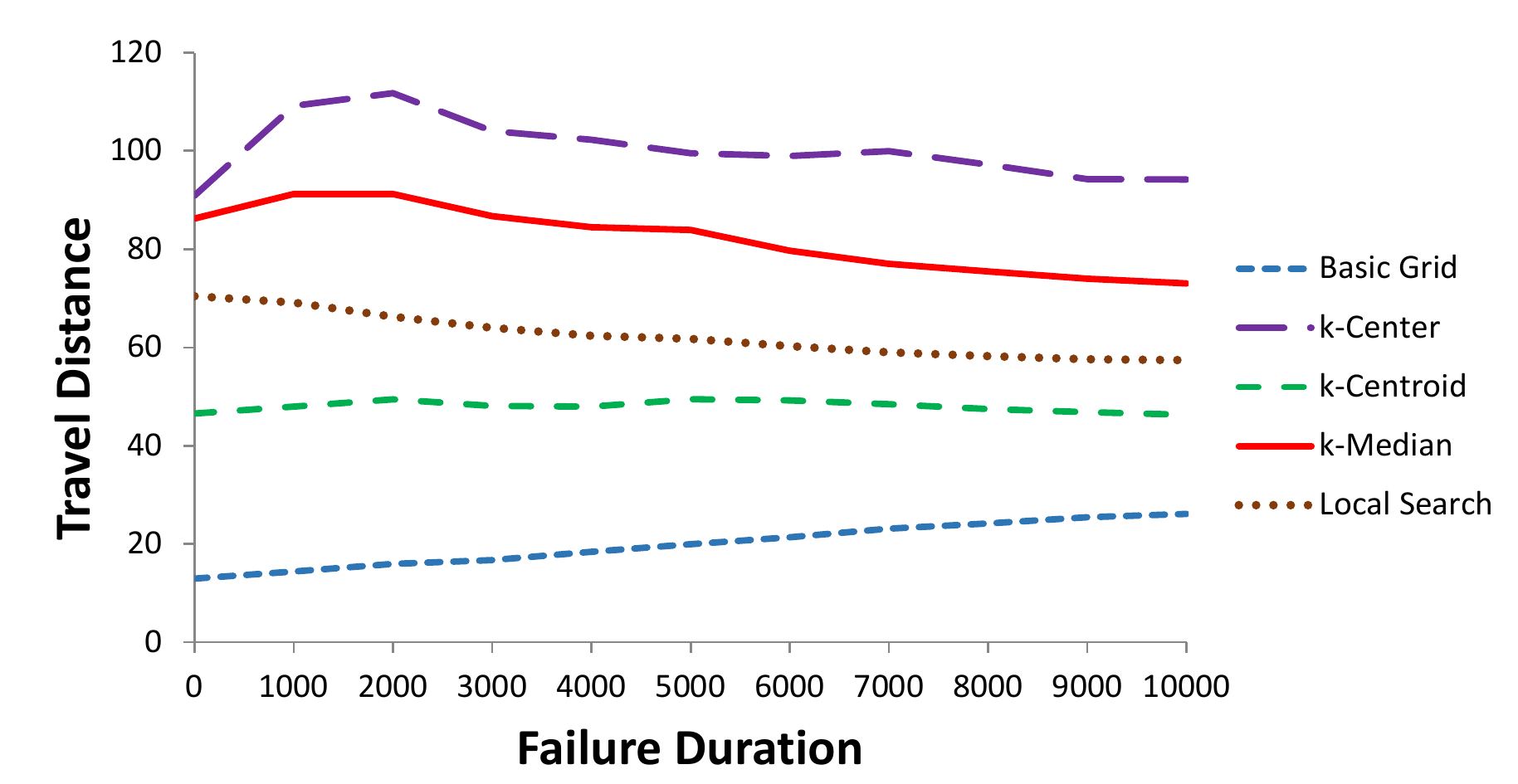}
    	 \caption{Average distance per mule as a factor of increasing failure durations.}
`	 \label{Fig:Figure2AM}
 \end{minipage}\hfill
	\vspace{-25pt}
\end{figure}

The results depicted in Figure~\ref{Fig:Figure1ADT} demonstrate the advantage of the $k$-Median and the $k$-Centroid algorithms in terms of average downtime.  The difference in the results of the $k$-Median algorithm to the Local Search algorithm and to the Basic Grid algorithm is statistically significant within $5\%$. The $k$-Center algorithm performs worse than the Basic Grid algorithm.

Figure~\ref{Fig:Figure2AM} presents a comparison between the algorithms in terms of the average distance traveled per mule. The results indicate that the $k$-Median and $k$-Center algorithms cause significantly more movement than the other algorithms. This is due to the fact that after every movement of a mule to a failure, a new deployment is calculated according to the occupied mules. As a result, all the available mules move with every failure as opposed to the Basic Grid algorithm where only one mule moves. The centroid-adjustment phase and local search create less movement since they only fine-tune the positions of the mules and in most cases only the mules that are very close to the one that moved are effected. In our experiments, usually after one or two rounds no mules move and only the nearest neighbors are effected. The reason that we see almost 10 times more movement in redeploying algorithms compared with the Basic Grid algorithm is due to the fact that there are 10 mules in this experiment.
Another interesting phenomena is that as failure durations increase,
the average movement in $k$-Median and $k$-Center decreases since there are less
un-occupied mules to reposition.


\begin{figure}
		\vspace{-15pt}
	\centering
	\begin{minipage}{0.45\textwidth}
		
		\centering
		\includegraphics[width=6cm]{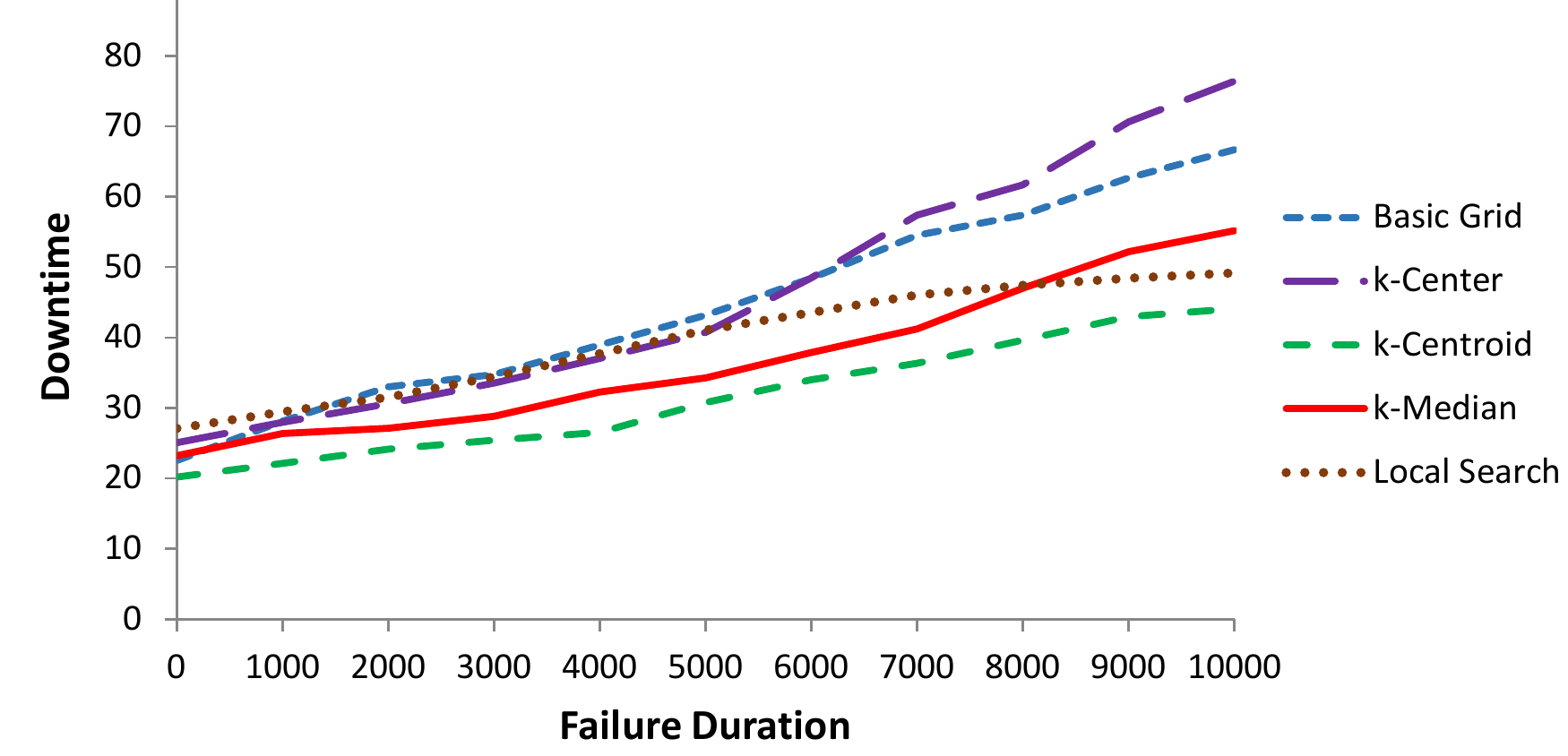}
	 \caption{Maximal downtime per failure as a factor of increasing failure durations.}
		\label{Fig:Figure3MDT}
		
	\end{minipage}\hfill
	\begin{minipage}{0.45\textwidth}
		\centering
		\includegraphics[width=6cm]{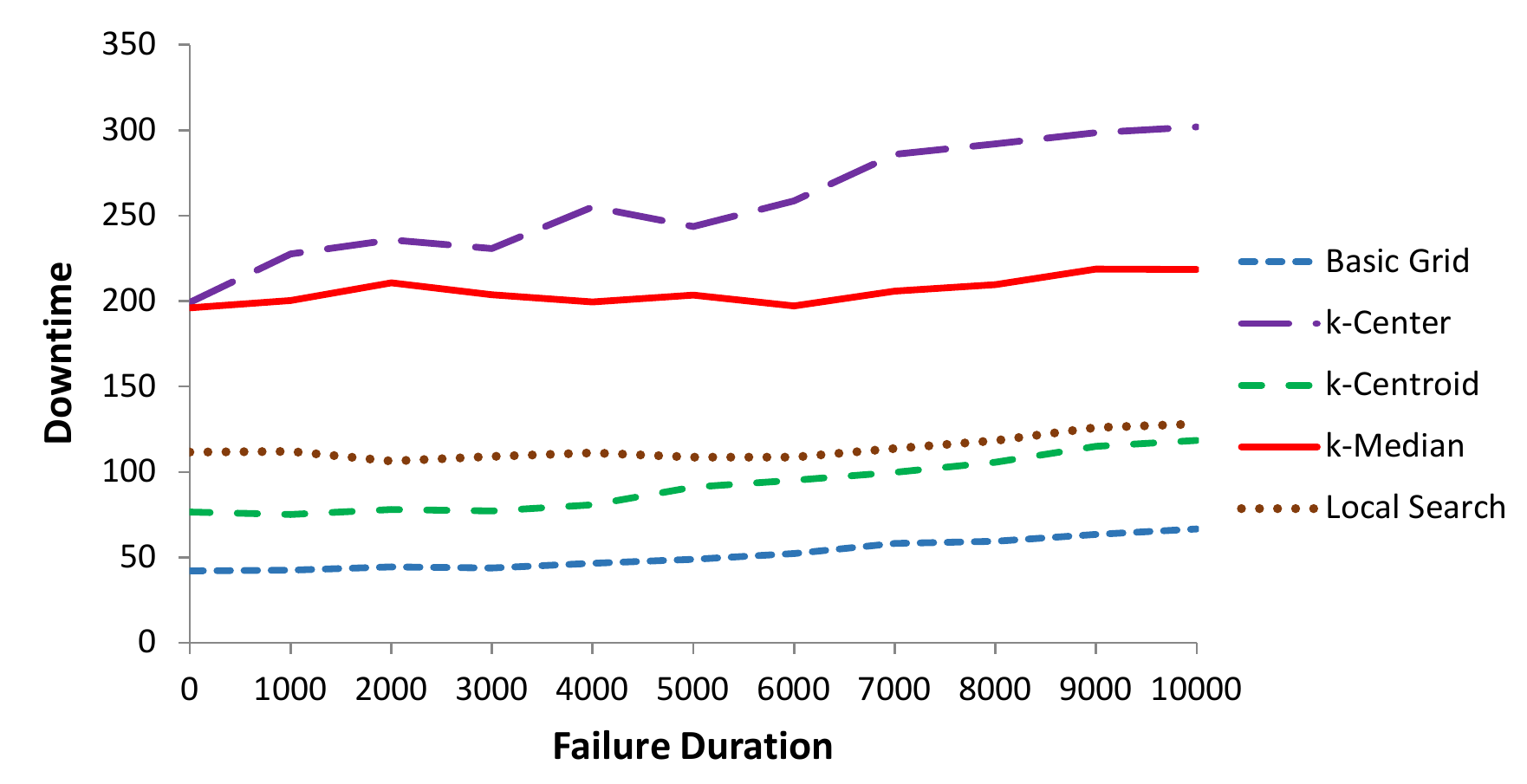}
  \caption{Maximal traveled distance of all the mules as a factor of increasing failure durations.}
		`	 \label{Fig:Figure4MM}
	\end{minipage}\hfill
	\vspace{-10pt}
\end{figure}

Figure~\ref{Fig:Figure3MDT} presents a comparison of algorithms in terms of the maximal downtime experienced by any node. The results show an advantage of the $k$-Centroid algorithms though the differences from it to the  $k$-Median algorithm are not statistically significant within $5\%$.

Figure~\ref{Fig:Figure4MM} presents a comparison between the algorithms in terms of the maximal distance traveled per mule. As in the average traveled distance, here too the results indicate that the $k$-Median and $k$-Center algorithms cause significantly more movement than the Basic Grid, Local Search and the $k$-Centroid algorithms.

\subsection{Non Uniform Failure Distribution}
\label{Sec:nonUniform}

In this case the failures were not generated randomly with equal
probability of any node failing. Instead, once a node fails, the
probability of failures in its vicinity is increased.

\begin{figure}
      \vspace{-10pt}
      \begin{center}
    \includegraphics[scale=0.45]{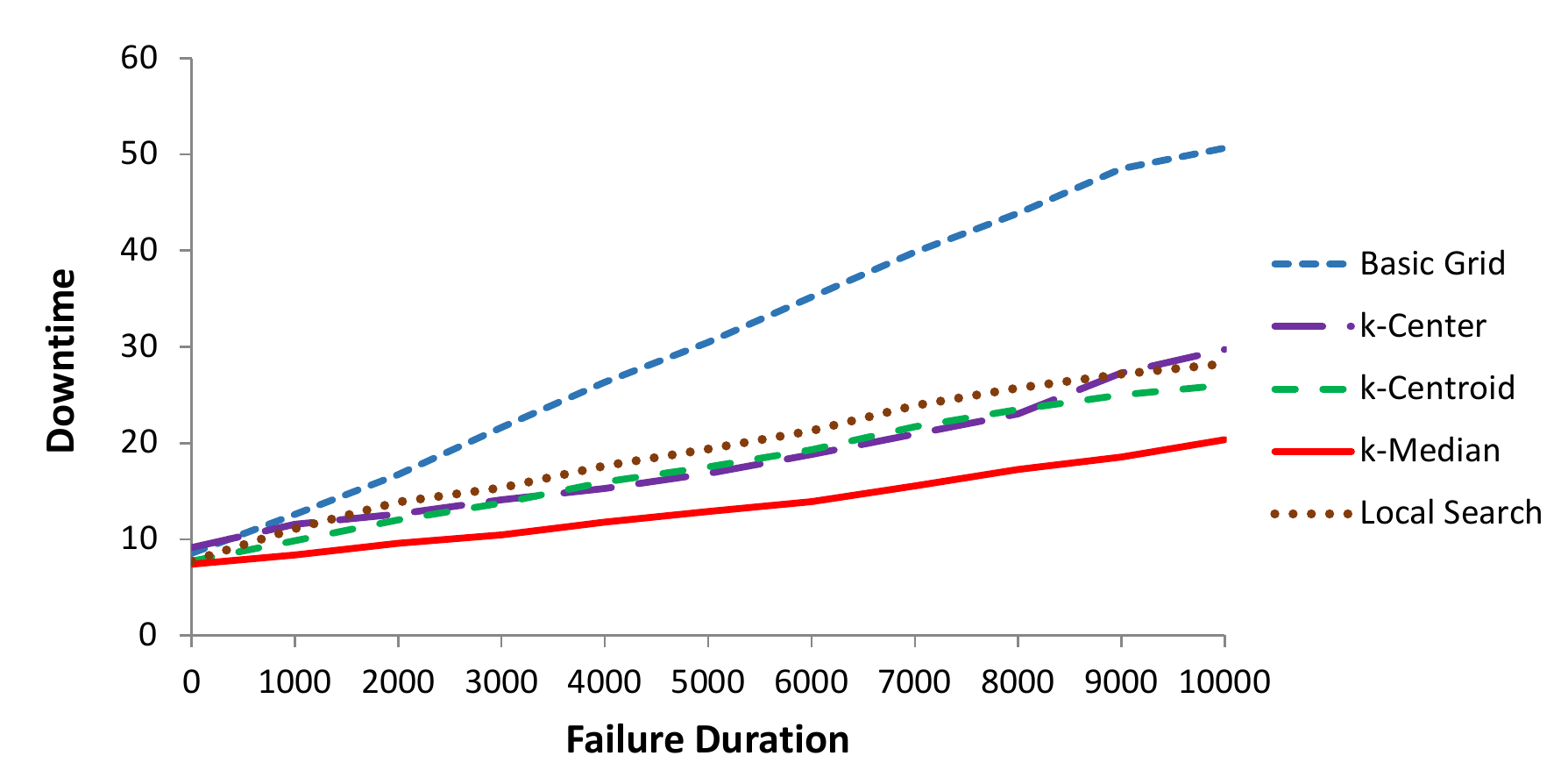}
           \vspace{-10pt}
    \caption{Average downtime with non-uniform failure distribution.}
    \label{Fig:FigureADNUD}
      \vspace{-20pt}
      \end{center}
\end{figure}

Figure~\ref{Fig:FigureADNUD} presents a comparison between the
algorithms in terms of the average downtime per failure. It is
interesting to see that unlike the uniform failure scenario, here
the $k$-Center algorithm performs very good and significantly better
than the Basic Grid algorithm. Here too the $k$-Median algorithm produces the best results, significantly better than the $k$-Centroid algorithm.


\section{Conclusion}
In this paper we proposed the use of facility location approximation algorithms for the coordination of a team of mobile agents charged with maintaining a WSN.
Specifically, we designed three algorithms based on approximation algorithms for the $k$-Center and $k$-Median problems and compared them to two baseline algorithms. The first, Basic Grid, is the best performing algorithm which does not use any redeployment techniques. The second, Local Search, enables to compare the algorithms to a heuristic local search approach. Our empirical results indicate that:
\begin{itemize}
    \item
    Redeployment using the $k$-Median heuristic (RGreedy) paired with a task allocation strategy that sends the closest-available mule to any failure, provides the best results in terms of minimizing the nodes' downtime.
    \item
    Algorithms that perform redeployment using either FF or RGreedy are less efficient in terms of mobility than using centroid adjustment or not redeploying at all.

    \item The $k$-Center algorithm performs poorly when failures are uniformly distributed but produces good results in non-uniform failure distribution in terms of downtime.

    \item
    Cooperative strategies that enable mules to tend to failed nodes outside of their Voronoi cells are more effective than non-cooperative ones and this advantage becomes more apparent with larger failure durations.

    \item
    Allocation of the closest-available mule to a failure is more effective than allocating the closest mule in terms of downtime and performs only slightly worse in terms of mobility. This advantage become apparent with larger failure durations.
\end{itemize}

A key contribution of this paper is the introduction of facility location approximation algorithms into the well studied AI problem of multi-agent coordination.

\section{Acknowledgments}
The research was been supported by the following sources:
Israel Science Foundation grant No. 1055/14 and grant No. 317/15, IBM Corporation, the Israeli Ministry of Economy and Industry, and the Helmsley Charitable Trust through the Agricultural, Biological and Cognitive Robotics Initiative of Ben-Gurion University of the Negev.

\bibliographystyle{abbrv}
\bibliography{references}
\end{document}